\documentclass[letterpaper]{article}
\usepackage{ijcai13}
\usepackage{graphicx}
\usepackage{algorithmic}
\usepackage{algorithm}
\usepackage{rotating}
\usepackage{amssymb}
\usepackage{xyling}

\usepackage{times}

\title{Recommender Systems for Configuration Knowledge Engineering \thanks{The work presented in this paper has been funded by the Austrian Research Promotion Agency (Project: ICONE (827587)).}}

\author{A. Felfernig, S. Reiterer, M. Stettinger, F. Reinfrank, M. Jeran, and G. Ninaus\\
Graz University of Technology\\
Inffeldgasse 16b, A-8010 Graz, Austria\\
\{felfernig,reiterer,stettinger,reinfrank,jeran,ninaus\}@ist.tugraz.at}

\begin{document}

\maketitle

\begin{abstract}
The knowledge engineering bottleneck is still a major challenge in configurator projects. In this paper we show how recommender systems can support knowledge base development and maintenance processes. We discuss a couple of scenarios for the application of recommender systems in knowledge engineering and report the results of  empirical studies which show the importance of user-centered configuration knowledge organization.
\end{abstract}

\section{Introduction}

Product  knowledge changes   frequently \cite{xcon}. Therefore,  it must be possible to conduct knowledge base development and maintenance operations efficiently. Since the early developments of configurator applications in the late 1970's and early 1980's \cite{McDermott82}, knowledge representations have been improved in terms of (1) \emph{model-based approaches} which allow a clear separation of domain knowledge and problem solving algorithms, (2) \emph{higher-level knowledge representations} which allow a component-oriented representation of configuration knowledge  (see, e.g., \cite{Stumptner98}), and (3) \emph{graphical knowledge representations} (e.g., \cite{felfernig00,Felfernig2001ConceptualModeling}) which allow  a  compact representation. In addition to new knowledge representations, intelligent \emph{diagnosis approaches} have been developed which help a knowledge engineer to identify and repair erroneous configuration knowledge \cite{paperjunker2004,paperfelfernig2004,paperfelfernig2009a,felfernigijcai2013}.

Due to diversification strategies of companies, product and service assortments are becoming increasingly large and complex \cite{Huffman1998}. The complexity of the underlying knowledge bases increases to the same extent which requires additional concepts that help a knowledge engineer to conduct knowledge base development and maintenance operations in an efficient fashion. Furthermore, knowledge bases are often developed by a group of persons with different knowledge, goals, and focuses with regard to  development and maintenance operations. This situation requires \emph{adaptive user interfaces} to be integrated into configuration knowledge engineering environments. Adaptive user interfaces for knowledge engineering have the potential  to effectively support engineers and domain experts in activities such as \emph{learning} (knowledge base understanding), \emph{finding} (the relevant items in the knowledge base), and \emph{testing \& debugging} (removing the source of faulty behavior).

In order to offer more adaptivity in configurator development environments, we propose the application of different types of recommendation technologies \cite{jan10} which proactively support domain experts and engineers when creating and adapting configuration knowledge. Such technologies should dispose of a basic understanding of cognitive processes when persons develop and maintain configuration knowledge bases. They should support functionalities such as \emph{recommending} relevant items (variables, component types, constraints, diagnoses, etc.)  and simultaneously \emph{omitting} specific items that are not relevant. Recommender systems  have the potential to provide such a support (see, e.g., \cite{paperrobillard2010}). 

There are three basic recommendation approaches. \emph{First}, \emph{collaborative filtering} \cite{paperkonstan1997} determines recommendations based on the preferences of \emph{nearest neighbors} (users with similar preferences compared to the current user). In this context, items are recommended to the current user which have received a positive rating by the nearest neighbors but are not known to the current user. \emph{Second}, \emph{content-based filtering} \cite{Pazzani1997} recommends items that are not known to the current user and are similar to items that have already been purchased by her/him. Similarity between items can be determined, for example, on the basis of the similarity of keywords used to describe the item. \emph{Third}, \emph{knowledge-based recommenders} recommend items  by using constraints or similarity metrics   \cite{paperburke2000,felfernigburke08}.

This paper is organized as follows. In Section \ref{Scenarios} we introduce  example  scenarios for the application of recommender technologies in knowledge engineering. Thereafter,    we report  results of related empirical studies (see Section \ref{Evaluation}). In Section \ref{RelatedWork} we provide a discussion of related work. Conclusions and a discussion of future research issues are given in Section \ref{Conclusions}.

\section{Recommenders for Knowledge Engineering} \label{Scenarios}

\textbf{Collaborative Recommendation of Constraints}. Collaborative filtering (CF) recommender systems have shown to be one of the best choices to achieve \emph{serendipity effects}, i.e., to be surprised (in a positive sense) by item recommendations one did not expect when starting the recommendation process. In situations were knowledge engineers do not know the configuration knowledge base very well, collaborative recommendations can be exploited to support a more focused analysis of the knowledge base. The availability of  navigation data from other knowledge engineers is the major precondition for determining recommendations with collaborative filtering. Table \ref{tab:cf_ constraints} shows an   example of navigation data that describes in which order knowledge engineers (users) accessed the  constraints of a knowledge base. For simplicity we assume that each of the users accessed each constraint (but in different order). Similar applications of collaborative filtering can be imagined for the recommendation of variables (or component types) and  instances  of a component catalog.

Table \ref{tab:cf_ constraints} stores the information in which order the constraints have been visited by knowledge engineers (users), for example, user $1$ analyzed the constraints in the order [$c_5,c_2,c_3,c_1,c_4,c_6$]. Let us assume that the current user has already visited the constraints $c_5$ and (then) $c_2$. The nearest neighbors of the current user (users with a similar navigation behavior) are the users $1$, $2$, and $4$. The majority of these users analyzed constraint $c_1$ in the third step -- this one will be recommended to the current user. Note that this recommendation approach is currently under evaluation, therefore no related empirical results will be reported in Section \ref{Evaluation}.

\begin{table} [ht]
\centering{}\begin{tabular}{|c|c|c|c|c|c|c|} 
\hline 
\emph{user} & \emph{$c_1$} & \emph{$c_2$} & \emph{$c_3$} & \emph{$c_4$} & \emph{$c_5$} & \emph{$c_6$}     \tabularnewline
\hline
\hline 
1 & 4 & 2  & 3  & 5  & 1 & 6  \tabularnewline
\hline 
2 & 3 & 2  & 5  & 6  & 1 & 4  \tabularnewline
\hline 
3 & 1 & 3  & 2  & 4  & 6 & 5  \tabularnewline
\hline 
4 & 3 & 2  & 4  & 5  & 1 & 6  \tabularnewline
\hline 
current  & ? & 2  & ?  & ?  & 1 & ?  \tabularnewline
\hline
\end{tabular} 
\caption{\label{tab:cf_ constraints} Recommending  constraints ($c_i$) with \emph{CF}.}
\end{table}

\textbf{Content-based Clustering of Constraints}. Another possibility to support knowledge engineers  is to cluster constraints with the goal to improve the overall clarity of the knowledge base. We will exemplify this on the basis of \emph{k-means clustering} \cite{book2005witten}. Following this approach, we have to generate $k$ initial \emph{centroids} which act as (first) representatives of future clusters. In the following, each object (in our case: constraint) is assigned to the group (cluster) with the closest (most similar) centroid. Thereafter, centroids are recalculated. In our case, a centroid is defined as the object with the highest overall similarity  to  the other objects in the cluster. The algorithm terminates if the centroids are \emph{stable} (do not change). \emph{k-means clustering} is guaranteed to terminate but is not necessarily optimal  since the  outcome depends on the initial centroids (\cite{book2005witten}).

For demonstration purposes we introduce the following simple configuration problem which is represented as a basic constraint satisfaction problem (CSP = (V, D, C)) where $V$ represents a set of variables \{$v_1, v_2, ..., v_5$\}, $D$ represents the set of corresponding domains ($dom(v_i) = \{1..5\}$), and $C$ represents the following set of constraints.

 \{$c_1: v_1=3 \rightarrow v_2 > 1,$
$ c_2: v_1 = 3 \land v_3 = 1,$
$  c_3: v_2 = 2 \rightarrow v_3 = 1,$
$  c_4: v_3 = 1 \rightarrow v_1 \neq 1,$
$ c_5: v_3 = 1 \rightarrow (v_4 = 2 \land v_1 > v_5),$
$ c_6: v_4 \geq 1 \rightarrow v_5 \leq 4,$
$ c_7: v_5 = 1 \rightarrow v_3 = 2 \lor v_3 = 3$\}. 

On the basis of this simple knowledge base, we can calculate the similarities between the individual constraints ($c_a, c_b$) by using Formula \ref{eq:similarity}. In this formula, $V =variables(c_a) \cup variables(c_b)$, $co-$$occurrence(v,c_a,c_b)$ = 1 if $v$ is contained in both constraints on the same position, $co-$$occurrence(v,c_a,c_b)$ = 0.5 if $v$ is contained in both constraints but on a different position, and $co-$$occurrence(v,c_a,c_b)$ = 0 of no co-occurrence exists. Note that this is \emph{one possible approach to similarity determination}. We also compared this approach with operator-based similarity and a random assignment of constraints to clusters.

\begin{equation} \label{eq:similarity}
sim(c_a, c_b) = \frac{\sum_{v \in V}   \mbox{\emph{co--occurrence}}(v,c_a,c_b)}{|V|}
\end{equation}

The similarities between the pairs of individual constraints are depicted in Table \ref{tab:constraintsimiliarty}. 

\begin{table} [ht]
\centering{}\begin{tabular}{|c|c|c|c|c|c|c|c|c|c|c|c|c|c|c|c|} 
\hline 
\emph{$c_i \in C$} & \emph{$c_1$} & \emph{$c_2$} & \emph{$c_3$} & \emph{$c_4$} & \emph{$c_5$} & \emph{$c_6$}   & \emph{$c_7$}  \tabularnewline
\hline
\hline 
$c_1$  & 1.0          &-        & -         & -        & -         & -       &  -  \tabularnewline
\hline 
$c_2$  & 0.33        &1.0    & -         & -        & -         & -       &  -  \tabularnewline
\hline 
$c_3$  & 0.16        &0.33  & 1.0     & -        & -         & -       &  -  \tabularnewline
\hline 
$c_4$  & 0.16        &0.5    & 0.16   & 1.0    & -         & -       &  -  \tabularnewline
\hline 
$c_5$  & 0.1          &0.25  & 0.1     & 0.37    & 1.0     & -       &  -  \tabularnewline
\hline 
$c_6$  & 0.0          &0.0    & 0.0     & 0.0    & 0.12   & 1.0   &  -  \tabularnewline
\hline 
$c_7$  & 0.0          &0.33  & 0.33   & 0.16  & 0.12 & 0.16 &  1.0  \tabularnewline
\hline 
\end{tabular} 
\caption{\label{tab:constraintsimiliarty} Similarities between individual constraints.}
\end{table}

On the basis of these individual similarities we are able to determine a set of corresponding clusters ($k = 2$). The determination of such clusters is exemplified in Table \ref{tab:constraintclustering}. First, we (randomly) select two constraints as initial cluster centers (centroids): $c_1$ and $c_5$ (denoted by $cs$). In iteration $2$ the center of cluster $1$ changes to $c_2$ and we have to re-calculate the cluster assignment. After this iteration, the assignment is stable, i.e., the cluster centers ($c_2$ and $c_5$) remain the same.

\begin{table} [ht]
\centering{}\begin{tabular}{|c|c|c|c|c|c|c|c|c|c|c|c|c|c|c|c|} 
\hline 
\emph{iteration} & \emph{$c_1$} & \emph{$c_2$} & \emph{$c_3$} & \emph{$c_4$} & \emph{$c_5$} & \emph{$c_6$}   & \emph{$c_7$}  \tabularnewline
\hline
\hline 
1  & $1 (cs)$          & $1 $        & 1         & 2        & $2 (cs)$         & $2$       &  $2$  \tabularnewline
\hline 
2  & $1 $          & $1 (cs)$        & 1         & 1        & $2 (cs)$         & $2$       &  $1$  \tabularnewline
\hline 
\end{tabular} 
\caption{\label{tab:constraintclustering} k-means clustering of $C = \{c_1, c_2, ..., c_7\}$.}
\end{table}

For the visualization of the constraints \{$c_1, c_2, ..., c_7$\} this means that the knowledge base would be presented in terms of two constraint groups: \{$c_1, c_2, c_3, c_4,c_7$\} and \{$c_5, c_6$\}.

\textbf{Knowledge-based Refactoring Recommendations}. The way in which semantics is expressed has an impact on the understandability of the knowledge base. For example, users need less time to understand the semantics of a knowledge base if implications are expressed in terms of $A \rightarrow B$ compared to the alternative representation of $\neg A \lor B$. Explicit knowledge about the cognitive complexity of constraint representations can be exploited to recommend structural and semantics-preserving adaptations of knowledge structures. Such recommendations are knowledge-based, since they are explicitly encoded in  refactoring rules.

\section{Empirical Evaluation} \label{Evaluation}
For the \emph{content-based clustering of constraints} and  \emph{knowledge-based refactoring recommendations} we  now  present the results of two  empirical studies. In the first study, we compared the applicability of three different clustering strategies with regard to knowledge engineering tasks (\emph{find a solution}, \emph{find a minimal conflict}) (see, e.g., \cite{paperjunker2004}).

\textbf{Study A: Clustering of Constraints}. For two different configuration knowledge bases ($kba_{1}, kba_{2}$) we conducted a study based on an within-subjects design (N=40). Each study participant  (students of computer science who visited a related course on knowledge engineering) had the task of (1) finding a solution (in $kba_1$) and (2) finding a minimal conflict (in $kba_2$).\footnote{We used these tasks to measure  knowledge  understanding. Further more differentiated tasks are within the scope of future work.} There were no time limits regarding task completion. Each student was assigned to one type of clustering (one out of variable-based similarity, operator-based similarity, and random clustering), i.e., we did not vary the type of clustering per student. The knowledge bases  ($kba_{1}, kba_{2}$) were defined as CSPs in a domain-independent fashion in order to avoid an additional cognitive complexity related to the understanding of a product domain. The basic properties of the used knowledge bases are summarized in Table \ref{tab:settingstudy1}.

\begin{table} [ht]
\centering{}\begin{tabular}{|c|c|c|c|c|c|c|c|c|c|c|c|c|c|c|c|} 
\hline 
Knowledge base & \#($v_i \in V$) & $v_i$ domain size & \#($c_i \in C$)     \tabularnewline
\hline
\hline 
$kba_1$  & 5          & 5 &    15       \tabularnewline
\hline 
$kba_2$  & 10          & 3 &    10       \tabularnewline
\hline 
\end{tabular} 
\caption{\label{tab:settingstudy1} Knowledge bases used in \emph{Study A}.}
\end{table}









The outcome of this experiment is shown in Table  \ref{tab:clusteringstudyerrors}. 

\begin{table} [ht]
\centering{}\begin{tabular}{|c|c|c|c|c|c|c|c|c|c|c|c|c|c|c|c|} 
\hline 
\emph{Grouping approach} & \emph{$kba_1$: SOL} & \emph{$kba_2$: CON} 
   \tabularnewline
\hline
\hline 
Similar variables & 21.43\% & 42.86\%    \tabularnewline
\hline 
Similar operators & 30.77\% & 53.85\%    \tabularnewline
\hline 
Random & 38.46\% & 76.92\%    \tabularnewline
\hline 
\end{tabular} 
\caption{\label{tab:clusteringstudyerrors} Error rates for completing the tasks \emph{find a solution (SOL)} and \emph{find a conflict (CON)} depending on clustering approach (variable-based, operator-based, or random).}
\end{table}

From the three compared approaches to the clustering of constraints in a configuration knowledge base, variable similarity based clustering clearly outperforms operator-based clustering and random clustering of  constraints.

\textbf{Study B: Cognitive Complexities}. There are different possibilities to represent equivalent semantics on the basis of a constraint, for example, the \emph{requires} relationship $X \rightarrow Y$ can be  represented in terms of $\neg X \lor Y$. The \emph{incompatibility} relationship $\neg(X \land Y)$ can be represented as $X \rightarrow \neg Y$. Table \ref{tab:CognitiveComplexity} depicts five different possibilities to express \emph{requires} and \emph{incompatibility} relationships. 

\begin{table} [ht]
\centering{}\begin{tabular}{|c|c|c|c|c|c|c|c|c|c|c|c|c|c|c|c|} 
\hline 
 $Requires$ & $Incompatibility$ 
   \tabularnewline
\hline
\hline 
 $X \rightarrow Y$ & $X \rightarrow \neg Y$    \tabularnewline
\hline 
 $\neg X \lor Y$  & $\neg X \lor \neg Y$    \tabularnewline
\hline 
 $\neg Y \rightarrow \neg X$ & $Y \rightarrow \neg X$    \tabularnewline
\hline 
 $\neg(X \land \neg Y)$ & $\neg (X \land Y)$    \tabularnewline
\hline 
$Y \leftarrow X$ & $\neg Y \leftarrow X$    \tabularnewline
\hline 
\end{tabular} 
\caption{\label{tab:CognitiveComplexity} Five different possibilities of representing \emph{requires} and \emph{incompatibility} relationships.}
\end{table}

Study B is  based on an within-subjects design (N=66) with two configuration knowledge bases. Knowledge base $kbb_1$ consisted of a set of  \emph{requires} constraints and $kbb_2$ consisted of a set of  \emph{incompatibility} constraints. Each study participant (again, computer science students who visited a related knowledge engineering course) had the task of finding a solution for the given CSP. Each participant was confronted with one version of $kbb_1$ and one version of $kbb_2$ conform the schema depicted in Table \ref{tab:CognitiveComplexity}. For example, if a student received the $X \rightarrow Y$ version of $kbb_1$ then she/he also received the $X \rightarrow \neg Y$ version of $kbb_2$. The knowledge bases $kbb_1$ and $kbb_2$ were (again) defined in a domain-independent fashion (see Study A). The basic properties of the used knowledge bases are summarized in Table \ref{tab:settingstudy2}.

\begin{table} [ht]
\centering{}\begin{tabular}{|c|c|c|c|c|c|c|c|c|c|c|c|c|c|c|c|} 
\hline 
Knowledge base & \#($v_i \in V$) & $v_i$ domain size & \#($c_i \in C$)     \tabularnewline
\hline
\hline 
$kbb_1$  & 5          & 5 &    7       \tabularnewline
\hline 
$kbb_2$  & 3          & 3 &    5       \tabularnewline
\hline 
\end{tabular} 
\caption{\label{tab:settingstudy2} Knowledge bases used in \emph{Study B}.}
\end{table}









The outcome of this experiment is shown in Table \ref{tab:resultcomplexityerrors}. 

\begin{table} [ht]
\centering{}\begin{tabular}{|c|c|c|c|c|c|c|c|c|c|c|c|c|c|c|c|} 
\hline 
 \emph{$kbb_1$: SOL} &  errors & \emph{$kbb_2$: SOL}  & errors
   \tabularnewline
\hline
\hline 
 $X \rightarrow Y$ & 21.43\% & $X \rightarrow \neg Y$  & 14.29\%  \tabularnewline
\hline 
 $\neg X \lor Y$  & 50.0\% & $\neg X \lor \neg Y$  &  34.62\%  \tabularnewline
\hline 
 $\neg Y \rightarrow \neg X$ & 96.43\% & $Y \rightarrow \neg X$  & 50.0\%   \tabularnewline
\hline 
 $\neg(X \land \neg Y)$ & 73.08\% & $\neg (X \land Y)$  &  42.31\% \tabularnewline
\hline 
$Y \leftarrow X$ & 25.0\% & $\neg Y \leftarrow X$  & 16.67\%   \tabularnewline
\hline 
\end{tabular} 
\caption{\label{tab:resultcomplexityerrors} Error rates in solution identification ($SOL$) depending on  constraint representation.}
\end{table}

A result of the study is that basic implications ($\rightarrow$) should be preferred to other representations in order to maximize understandability. The only type of knowledge representation with a similar performance is the reverse implication, however, when comparing both alternatives, the standard implication seems to be the better choice.

\section{Related Work} \label{RelatedWork}

There is a long history of research on the improvement of knowledge engineering processes. Early research focused on model-based knowledge representations that allowed a  separation of domain and problem solving knowledge. An example of such a representation are constraint technologies  which became extremely popular as a technological basis for industrial applications \cite{freuder97}. In a next step, graphical knowledge representations \cite{felfernig00} and intelligent techniques for knowledge base testing and debugging have been developed \cite{paperfelfernig2004}. The need of an intuitive access to a corpus of software artifacts is also one of the major requirements for \emph{software comprehension} \cite{Storey2006}. In this context, recommender systems \cite{jan10} have already been identified as a valuable means to provide intelligent support for the navigation in large and complex software spaces (see, e.g.,  \cite{paperrobillard2010}). The application of recommendation technologies for supporting knowledge engineering processes is a new research area. Research contributions in this field have the potential to significantly improve the overall quality of knowledge engineering processes. In \cite{paperfelfernigpum2010} basic  knowledge representations are compared, for example, the use of $\rightarrow$ to represent an implication vs. the use of $\neg$ and $\lor$. This work is an important step towards a discipline of \emph{empirical knowledge engineering} with a clear focus on usability aspects and cognitive efforts needed to complete knowledge engineering tasks.  The work presented in this paper is a continuation of the work of \cite{paperfelfernigpum2010}. It takes a more detailed look at different alternative representations of \emph{requires} and \emph{incompatibility} relationships and introduces a new concepts related to the content-based clustering of constraints.

\section{Conclusions} \label{Conclusions}
In this paper we showed how recommenders can be exploited to support knowledge engineering tasks. Examples are collaborative filtering of constraint sets,  clustering of constraints, and  knowledge-based recommendation of refactoring operations. Future work will include the development of further recommendation algorithms, for example, the inclusion of content-based filtering and further clustering algorithms as well as further empirical studies with  more differentiated maintenance tasks. Finally, we will  focus on an in-depth analysis of existing research in the area of cognition psychology which can further advance the state of the art in (configuration) knowledge engineering.

\bibliographystyle{named}
\bibliography{references}

\end{document}